\documentclass[10pt]{article}
\usepackage{graphicx}
\usepackage{amsmath}
\usepackage{amssymb}
\usepackage{multirow}
\usepackage{caption2}
\usepackage[figuresright]{rotating}
\begin{document}
\bibliographystyle{prsty}
\begin{center}
{\large {\bf \sc{Strong decays of the $P_{cs}(4338)$ and its high isospin cousin via the QCD sum rules}}} \\[2mm]
Xiu-Wu Wang, Zhi-Gang Wang\footnote{E-mail: zgwang@aliyun.com.}\\
 Department of Physics, North China Electric Power University, Baoding 071003, P. R. China\\
\end{center}

\begin{abstract}
In the present work, the strong decays of the newly observed $P_{cs}(4338)$ as well as its high isospin cousin $P_{cs}(4460)$ are studied via the QCD sum rules. According to conservation of isospin, spin  and parity, the hadronic coupling constants in four decay channels are obtained, then the partial decay widths are obtained. The total width of the $P_{cs}(4338)$ coincides with the experimental data nicely, while the predictions for the $P_{cs}(4460)$ can be testified in the future experiment, and shed light on the nature of the $P_{cs}(4338)$.
\end{abstract}



\section{Introduction}
In recent years, several $P_c$ and $P_{cs}$ exotic states, such as the $P_c(4312)$, $P_c(4380)$, $P_c(4440)$, $P_c(4457)$, $P_c(4337)$, $P_{cs}(4459)$ and $P_{cs}(4338)$, were observed by the LHCb collaboration \cite{RAaij1,RAaij2,LHCb-Pcs4459-2012,LHCb-Pc4337-2108,LHCb-Pcs4338}, they are hidden-charm pentaquark (molecule) candidates with or without strangeness. Except for the $P_c(4337)$, the $P_c$ and $P_{cs}$ exotic states lie near the thresholds of the  $\Sigma_c^{(*)}\bar{D}^{(*)}$ and $\Xi_c\bar{D}^{(*)}$ pairs, respectively, it is natural to consider them as the meson-baryon molecules or color singlet-singlet type pentaquark states. In the present study, we will focus on the exotic $P_{cs}(4338)$ observed in the $J/\psi\Lambda$ channel \cite{LHCb-Pcs4338}, its measured  Breit-Wigner mass and width are $4338.2\pm0.7\pm0.4\,\rm{MeV}$ and $7.0\pm1.2\pm1.3\,\rm{MeV}$, respectively, and the preferred spin-parity is $J^P={\frac{1}{2}}^-$ \cite{LHCb-Pcs4338}, its discovery is in line with observation of the $P_c^\pm(4337)$ in the $J/\psi p$ and $J/\psi \bar{p}$ channels \cite{LHCb-Pc4337-2108}.

In Ref.\cite{MJYan-Pcs4338-mole}, the $P_{cs}(4338)$ is considered as the $\Xi_c\bar{D}$ molecule  via the effective field theory.
In Ref.\cite{PGOrtega-Pcs4338-mole},
 the mass and width of the $P_{cs}(4338)$ are studied by assigning it as the meson-baryon molecule with the $IJ^P=0\frac{1}{2}^-$ based on the constituent quark model, and  partial widths of its strong decays are obtained in details. In Ref.\cite{JTZhu-Pcs4338-mole}, the $P_{cs}(4338)$ is interpreted as the $\Xi_c\bar{D}$ molecular state  with the $J^P=\frac{1}{2}^-$ via the  quasipotential Bethe-Salpeter equation, and new structures are predicted. In Ref.\cite{KAzizi-Pcs4338-mole}, the mass and decays of the $P_{cs}(4338)$ are studied with the QCD sum rules by considering it as the $\Xi_c\bar{D}$ molecule with the spin-parity $J^P=\frac{1}{2}^-$ . For some other interesting works focusing on the $P_{cs}(4338)$, one can consult Refs.\cite{LMeng-Pcs4338,XWWang-Pcs4338-mole,UOzdem-Pcs4338-magnetic,SXNakamura-Pcs4338,FLWang-PcPcs,BWang-PcPcs,Burns-Pcs4338,Giachino-Pcs4338,KChen-Pcs4338,ZYYang-Pcs4338,AFeijoo-Pcs4338,EYParyev-Pcs4338,GYang-Pcs4338,SXNakamura2-Pcs4338,QWu-Pcs4338}. Along with the popular  acceptance of the meson-baryon molecule assignment for the $P_{cs}(4338)$, debates do exist about its nature, for example,
 in Refs.\cite{SYLi-Pcs4338-compact,WXZhang-Pcs4338-compact}, the $P_{cs}(4338)$ is assigned as the compact pentaquark state.

For our research work on the $P_{cs}(4338)$, we consider that it has the definite  isospin, spin and parity $IJ^P=0\frac{1}{2}^-$ and identify it as the $\Xi_c\bar{D}$ hadronic molecule via the QCD sum rules \cite{XWWang-Pcs4338-mole}, moreover, our calculations show that there maybe exist a high isospin cousin $P_{cs}(4460)$, which is assigned as the $\Xi_c\bar{D}$ resonant state. In Refs.\cite{WangXW-SCPMA,WangXW-PentaMole-s}, we distinguish the isospin, spin and parity, study the mass spectrum of the hidden-charm pentaquark molecular states without strangeness and with strangeness in a systematic way, and make possible assignments of the existing pentaquark candidates and predict many new exotic states. If those predicted states could be observed in the future experiment, it would testify our interpretation of the nature of the $P_{cs}(4338)$, etc.
In Ref.\cite{XWWang-Pc4312-decay}, we study  the strong decays of the pentaquark molecule candidate  $P_c(4312)$ and its higher isospin cousin $P_c(4330)$ with the QCD sum rules to examine their nature.
Now we study the strong decays of the $P_{cs}(4338)$ and $P_{cs}(4460)$ in detail to provide some useful information for the future high energy experiment.

This paper is arranged as follows, the QCD sum rules for the strong decays are derived in Section 2,  the QCD sum rules for the $\Lambda$ and $\Sigma$ baryons are studied in Section 3, numerical calculations and discussions are presented in Section 4, and Section 5 is reserved for the conclusions.
\section{QCD sum rules for the strong decays of the $P_{cs}(4338)$ and $P_{cs}(4460)$}
In Ref.\cite{XWWang-Pcs4338-mole}, the quantum numbers $IJ^P$ of the exotic pentaquark molecular states with strangeness $P_{cs}(4338)$ and $P_{cs}(4460)$ are assigned as $0\frac{1}{2}^-$ and $1\frac{1}{2}^-$, respectively, the currents $J_{P}(x)$ and $J_{P^{\prime}}(x)$ are applied to interpolate the $P_{cs}(4338)$ and $P_{cs}(4460)$, respectively,
\begin{eqnarray}
J_{P}(x)&=&\frac{1}{\sqrt{2}}\varepsilon^{ijk}\left[u^{iT}(x)C\gamma_5s^j(x)c^{k}(x)\bar{c}(x)\textsf{i}\gamma_5d(x)-d^{iT}(x)C\gamma_5s^j(x)c^k(x)\bar{c}(x)\textsf{i}\gamma_5u(x)\right] \, , \nonumber\\
J_{P^{\prime}}(x)&=&\frac{1}{\sqrt{2}}\varepsilon^{ijk}\left[u^{iT}(x)C\gamma_5s^j(x)c^{k}(x)\bar{c}(x)\textsf{i}\gamma_5d(x)+d^{iT}(x)C\gamma_5s^j(x)c^k(x)\bar{c}(x)\textsf{i}\gamma_5u(x)\right] \, ,
\end{eqnarray}
where $\textsf{i}^2=-1$, $i$, $j$, $k$ are the color indices, the $C$ represents the charge conjugation matrix. It is worth mentioning that both the currents $J_{P}(x)$ and $J_{P^{\prime}}(x)$ are the color-singlet-color-singlet currents,  which are applied to study the related molecular states in the framework of the QCD sum rules. We are also interested in the argument that the $P_{cs}(4338)$ may be a compact pentaquark state \cite{SYLi-Pcs4338-compact,WXZhang-Pcs4338-compact}, furthermore, for the $P_{cs}(4459)$, we applied both the scalar-diquark-scalar-diquark-antiquark type current \cite{ZGWang-Pcs4459-pentaquark} and color singlet-singlet currents \cite{XWWang-Pcs4338-mole,WangXW-PentaMole-s} to interpolate this exotic state, similar works have been done for the $P_c(4312)$, $P_c(4440)$ and $P_c(4457)$ \cite{WangXW-SCPMA,ZGWang-Pc4312etc-pentaquark,ZGWang-Pc4450-pentaquark}. It will be surely meaningful to construct some new five-quark currents to explore the possible compact pentaquark candidate $P_{cs}(4338)$ in our future work.

At present, we consider conservation of the $IJ^P$ in the strong decays, and study the typical  decay channels,
\begin{eqnarray}
P_{cs}(4338)&\rightarrow&\eta_c+\Lambda\, , \nonumber\\
P_{cs}(4338)&\rightarrow&J/\psi+\Lambda\, , \nonumber\\
P_{cs}(4460)&\rightarrow&\eta_c+\Sigma\, , \nonumber\\
P_{cs}(4460)&\rightarrow&J/\psi+\Sigma\, ,
\end{eqnarray}
where the $IJ^P$ of the $\eta_c$, $J/\psi$, $\Lambda$ and $\Sigma$ are $00^-$, $01^-$, $0\frac{1}{2}^+$ and $1\frac{1}{2}^+$, respectively. The interpolating currents of those mesons and baryons are written as,
\begin{eqnarray}\label{Current}
J_{\eta_c}(x)&=&\bar{c}(x)\textsf{i}\gamma_5c(x)\, , \nonumber\\
J_{J/\psi,\mu}(x)&=&\bar{c}(x)\gamma_{\mu}c(x)\, , \nonumber\\
J_{\Lambda}(x)&=&\sqrt{\frac{2}{3}}\varepsilon^{ijk}\left[u^{iT}(x)C\gamma_{\alpha}s^j(x)\gamma_5\gamma^{\alpha}d^k(x)-d^{iT}(x)C\gamma_{\alpha}s^j(x)\gamma_5\gamma^{\alpha}u^k(x)\right]\, , \nonumber\\
J_{\Sigma}(x)&=&\varepsilon^{ijk}u^{iT}(x)C\gamma_{\alpha}d^{j}(x)\gamma^{\alpha}\gamma_5s^{k}(x)\, .
\end{eqnarray}
Now the three-point correlation functions for those  decay channels can be written as,
\begin{eqnarray}\label{CF-Pi}
\Pi_P(p,q) &=& \textsf{i}^2 \int d^4xd^4y e^{\textsf{i}p\cdot x}e^{\textsf{i}q\cdot y} \langle 0|\texttt{T} \left\{ J_{\eta_c}(x)J_{\Lambda}(y) \bar{J}_{P}(0) \right\}|0\rangle\, ,\\
\Pi_{P,\mu}(p,q) &=& \textsf{i}^2 \int d^4xd^4y e^{\textsf{i}p\cdot x}e^{\textsf{i}q\cdot y} \langle 0|\texttt{T} \left\{ J_{J/\psi,\mu}(x)J_{\Lambda}(y) \bar{J}_{P}(0) \right\}|0\rangle\, ,\\
\Pi_{P^{\prime}}(p,q) &=& \textsf{i}^2 \int d^4xd^4y e^{\textsf{i}p\cdot x}e^{\textsf{i}q\cdot y} \langle 0|\texttt{T} \left\{ J_{\eta_c}(x)J_{\Sigma}(y) \bar{J}_{P'}(0) \right\}|0\rangle\, ,\\
\Pi_{P^{\prime},\mu}(p,q) &=& \textsf{i}^2 \int d^4xd^4y e^{\textsf{i}p\cdot x}e^{\textsf{i}q\cdot y} \langle 0|\texttt{T} \left\{ J_{J/\psi,\mu}(x)J_{\Sigma}(y) \bar{J}_{P'}(0) \right\}|0\rangle\, ,
\end{eqnarray}
where the $\texttt{T}$ is the time-order operator. At the hadronic sides, a complete set of intermediate hadron states which have the same quantum numbers $IJ^P$  as the corresponding currents are inserted \cite{SVZ1,SVZ2,Reinders}, those correlation functions are shown as follows after the contributions of the ground states being isolated,
\begin{eqnarray}
\Pi_P(p,q) &=& \frac{f_{\eta_c}m_{\eta_c}^2}{2m_c}\lambda_{\Lambda}\lambda_{P} g_{\eta\Lambda} \frac{\left(\!\not\!{q}+m_{\Lambda}\right)\left(\!\not\!{p'}+m_{P}\right)}
{\left(m^2_{P}-p'^2\right)\left(m^2_{\eta_c}-p^2\right)\left(m^2_{\Lambda}-q^2\right)}+\cdots\, ,
\end{eqnarray}
\begin{eqnarray}
\Pi_{P,\mu}(p,q)&=&f_{J/\psi}m_{J/\psi}\lambda_{\Lambda}\lambda_{P}\frac{-\textsf{i}}{\left(m^2_{P}-p'^2\right)\left(m^2_{J/\psi}-p^2\right)\left(m^2_{\Lambda}-q^2\right)}\left( -g_{\mu\alpha}+\frac{p_{\alpha}p_{\mu}}{p^2}\right)\nonumber\\
&&\left(\!\not\!{q}+m_{\Lambda}\right)\left(g_{J/\psi\Lambda,V}\gamma^{\alpha}-\frac{\textsf{i}g_{J/\psi\Lambda,T}}{m_{P}+m_{\Lambda}}\sigma^{\alpha\beta}p_{\beta}\right)
\gamma_5\left(\!\not\!{p'}+m_{P}\right)+\cdots\, ,
\end{eqnarray}
\begin{eqnarray}
\Pi_{P^{\prime}}(p,q) &=& \frac{f_{\eta_c}m_{\eta_c}^2}{2m_c}\lambda_{\Sigma}\lambda_{P^{\prime}} g_{\eta\Sigma} \frac{\left(\!\not\!{q}+m_{\Sigma}\right)\left(\!\not\!{p'}+m_{P^{\prime}}\right)}
{\left(m^2_{P^{\prime}}-p'^2\right)\left(m^2_{\eta_c}-p^2\right)\left(m^2_{\Sigma}-q^2\right)}+\cdots\, ,
\end{eqnarray}
\begin{eqnarray}
\Pi_{P^{\prime},\mu}(p,q)&=&f_{J/\psi}m_{J/\psi}\lambda_{\Sigma}\lambda_{P^{\prime}}\frac{-\textsf{i}}{\left(m^2_{P^{\prime}}-p'^2\right)\left(m^2_{J/\psi}-p^2\right)\left(m^2_{\Sigma}-q^2\right)}\left( -g_{\mu\alpha}+\frac{p_{\alpha}p_{\mu}}{p^2}\right)\nonumber\\
&&\left(\!\not\!{q}+m_{\Sigma}\right)\left(g_{J/\psi\Sigma,V}\gamma^{\alpha}-\frac{\textsf{i}g_{J/\psi\Sigma,T}}{m_{P^{\prime}}+m_{\Sigma}}\sigma^{\alpha\beta}p_{\beta}\right)
\gamma_5\left(\!\not\!{p'}+m_{P^{\prime}}\right)+\cdots\, ,
\end{eqnarray}
where the $g_{\eta\Lambda}$, $g_{\eta\Sigma}$, $g_{J/\psi\Lambda,V}$, $g_{J/\psi\Lambda,T}$, $g_{J/\psi\Sigma,V}$ and $g_{J/\psi\Sigma,T}$ are the hadronic coupling constants, the $\lambda_{P}$, $\lambda_{P'}$, $\lambda_{\Lambda}$ and $\lambda_{\Sigma}$ are the pole residues of the $P_{cs}(4338)$, $P_{cs}(4460)$, $\Lambda$ and $\Sigma$, respectively, the $f_{\eta_c}$ and $f_{J/\psi}$ are the decay constants of the mesons $\eta_c$ and $J/\psi$, respectively, moreover, those constants satisfy the following definitions,
\begin{eqnarray}
\langle 0| J_{P}(0)|\mathcal{P}_{P}(p')\rangle &=& \lambda_{P}U_{P}(p')\,,\nonumber  \\
\langle 0| J_{P'}(0)|\mathcal{P}_{P'}(p')\rangle &=& \lambda_{P'}U_{P'}(p')\,,\nonumber  \\
\langle 0| J_{\Lambda}(0)|\Lambda(q)\rangle &=& \lambda_{\Lambda}U_{\Lambda}(q)\,,\nonumber \\
\langle 0| J_{\Sigma}(0)|\Sigma(q)\rangle &=& \lambda_{\Sigma}U_{\Sigma}(q)\,,\nonumber \\
\langle 0| J_{J/\psi,\mu}(0)|J/\psi(p)\rangle &=& f_{J/\psi}m_{J/\psi}\varepsilon_{\mu}\,,\nonumber \\
\langle 0| J_{\eta_c}(0)|\eta_{c}(p)\rangle &=& \frac{f_{\eta_c}m_{\eta_c}^2}{2m_c}\,,
\end{eqnarray}
\begin{eqnarray}\label{gP-gPprime}
\langle \eta_c(p)\Lambda(q)|\mathcal{P}_{P}(p')\rangle &=& \textsf{i}g_{\eta\Lambda}\overline{U}_{\Lambda}(q)U_{P}(p')\,,\nonumber \\
\langle J/\psi(p)\Lambda(q)|\mathcal{P}_{P}(p')\rangle &=& \overline{U}_{\Lambda}(q)\varepsilon_{\alpha}^*\left(g_{J/\psi\Lambda,V}\gamma^{\alpha}-\textsf{i}\frac{g_{J/\psi\Lambda,T}}{m_{P}+m_{\Lambda}}
\sigma^{\alpha\beta}p_{\beta}\right)\gamma_5U_{P}(p')\,,\nonumber \\
\langle \eta_c(p)\Sigma(q)|\mathcal{P}_{P^{\prime}}(p')\rangle &=& \textsf{i}g_{\eta\Sigma}\overline{U}_{\Sigma}(q)U_{P^{\prime}}(p')\,,\nonumber \\
\langle J/\psi(p)\Sigma(q)|\mathcal{P}_{P^{\prime}}(p')\rangle &=& \overline{U}_{\Sigma}(q)\varepsilon_{\alpha}^*\left(g_{J/\psi\Sigma,V}\gamma^{\alpha}-\textsf{i}\frac{g_{J/\psi\Sigma,T}}{m_{P^{\prime}}+m_{\Sigma}}
\sigma^{\alpha\beta}p_{\beta}\right)\gamma_5U_{P^{\prime}}(p')\,,
\end{eqnarray}
where the $|\mathcal{P}_{P}\rangle$, $|\mathcal{P}_{P'}\rangle$, $|\Lambda\rangle$, $|\Sigma\rangle$, $|\eta_{c}\rangle$ and $|J/\psi\rangle$ represent the ground states $P_{cs}(4338)$, $P_{cs}(4460)$, $\Lambda$, $\Sigma$, $\eta_{c}$ and $J/\psi$, respectively, the $U_{P}$, $U_{P^{\prime}}$, $U_{\Lambda}$ and $U_{\Sigma}$ are the Dirac spinors, the $\varepsilon_{\mu}$ stands for the polarization vector of the $J/\psi$ meson, it satisfies the formula  $\sum \varepsilon_{\mu}\varepsilon^*_{\alpha}=-g_{\mu\alpha}+\frac{p_{\mu}p_{\alpha}}{p^2}$.

It is reasonable to suppose that the hadronic sides $\Pi_{H}(p,q)$ and QCD sides $\Pi_{QCD}(p,q)$ of the correlation functions should match with each other \cite{XWWang-Pc4312-decay,Decay-mole-WZG-WX,WZG-Pc4312-decay-tetra},
\begin{eqnarray}
Tr[\Pi_{H}(p,q)\cdot\Gamma]&=&Tr[\Pi_{ QCD}(p,q)\cdot\Gamma]\,,
\end{eqnarray}
where the $\Gamma$ is any matrix in the Dirac spinor space. In the present study, the $\Gamma$ are chosen as $\sigma_{\mu\nu}$ and $\gamma_{\mu}$ for both the $\Pi_P(p,q)$ and $\Pi_{P^{\prime}}(p,q)$, when setting $\Gamma=\sigma_{\mu\nu}$, the tensor structure is chosen as $p_{\mu}q_{\nu}-q_{\mu}p_{\nu}$, as for $\Gamma=\gamma_{\mu}$, the structure $q_\mu$ is picked out. For the correlation functions $\Pi_{P,\mu}(p,q)$ and $\Pi_{P^{\prime},\mu}(p,q)$, the $\Gamma$ are chosen as $\gamma_5\!\not\!{z}$ and $\gamma_5$, then the resulting tensor structures are $q_\mu p\cdot z$ and $q_\mu$, respectively. For clarity, the chosen tensor structures are expressed as,
\begin{eqnarray}
\frac{1}{4}Tr[\Pi_{P}(p,q)\sigma_{\mu\nu}] &=& \Pi_a(p'^2,p^2,q^2)\textsf{i}(p_\mu q_\nu-q_\mu p_\nu)+\cdots\,,\nonumber \\
\frac{1}{4}Tr[\Pi_{P}(p,q)\textsf{i}\gamma_{\mu}] &=& \Pi_b(p'^2,p^2,q^2)\textsf{i}q_\mu+\cdots\,, \nonumber \\
\frac{1}{4}Tr[\Pi_{P,\mu}(p,q)\gamma_{5}\!\not\!{z}] &=& \Pi_c(p'^2,p^2,q^2)\textsf{i}q_\mu p\cdot z+\cdots\,, \nonumber\\
\frac{1}{4}Tr[\Pi_{P,\mu}(p,q)\gamma_{5}] &=& \Pi_d(p'^2,p^2,q^2)\textsf{i}q_\mu +\cdots\,,\nonumber \\
\frac{1}{4}Tr[\Pi_{P^{\prime}}(p,q)\sigma_{\mu\nu}] &=& \Pi_e(p'^2,p^2,q^2)\textsf{i}(p_\mu q_\nu-q_\mu p_\nu)+\cdots\,,\nonumber \\
\frac{1}{4}Tr[\Pi_{P^{\prime}}(p,q)\textsf{i}\gamma_{\mu}] &=& \Pi_f(p'^2,p^2,q^2)\textsf{i}q_\mu+\cdots\,, \nonumber \\
\frac{1}{4}Tr[\Pi_{P^{\prime},\mu}(p,q)\gamma_{5}\!\not\!{z}] &=& \Pi_g(p'^2,p^2,q^2)\textsf{i}q_\mu p\cdot z+\cdots\,, \nonumber\\
\frac{1}{4}Tr[\Pi_{P^{\prime},\mu}(p,q)\gamma_{5}] &=& \Pi_h(p'^2,p^2,q^2)\textsf{i}q_\mu +\cdots\,.
\end{eqnarray}

At the QCD sides, all the quark fields are contracted via the Wick theorem, and then the operator product expansions are performed, for the analytical calculations  of the quark fields, the integrals of the light and heavy quarks are solved in the coordinate space and momentum space, respectively \cite{Decay-mole-WZG-WX,WZG-Pc4312-decay-tetra}. Since the relation $p'=p+q$ holds for all the two-body decay channels considered in the present study, the $p'^2$ is set as $\xi p^2$, where the $\xi$ is a constant relied on the particular  decay channel \cite{XWWang-Pc4312-decay}, taking the decay channel $P_{cs}(4338)\rightarrow\eta_c+\Lambda$ for example, $\xi= \frac{m_{\Lambda}^2}{m_{\eta_c^2}}+1$. Following  the rigorous quark-hadron duality below the continuum thresholds \cite{WZG-ZJX-Zc-Decay,WZG-Y4660-Decay,WZG-X4140-decay,WZG-X4274-decay,WZG-Z4600-decay}, double Borel transformations are applied, then the QCD sum rules for the hadronic coupling constants are derived as,
\begin{eqnarray}\label{QCDSG-G-i}
&& \frac{f_{\eta_c}m_{\eta_c}^2\lambda_{\Lambda}\lambda_{P} g_{\eta\Lambda,a}}{2m_c \xi}\frac{1}{\frac{m_{P}^2}{\xi}-m_{\eta_c}^2}\left\{ {\rm exp} \left( -\frac{m_{\eta_c}^2}{T_1^2} \right)-{\rm exp} \left( -\frac{m_{P}^2}{\xi T_1^2} \right) \right\}{\rm exp} \left( -\frac{m_{\Lambda}^2}{T_2^2} \right) \nonumber \\
&& +C_a {\rm exp} \left( -\frac{m_{\eta_c}^2}{T_1^2}-\frac{m_{\Lambda}^2}{T_2^2} \right)=\int_{4m_c^2}^{s_{\eta_c}^0}ds\int_0^{s_{\Lambda}^0}du\, \rho_a(s,u){\rm exp}\left( -\frac{s}{T_1^2}-\frac{u}{T_2^2} \right)\,,
\end{eqnarray}
\begin{eqnarray}
&& \frac{f_{\eta_c}m_{\eta_c}^2\lambda_{\Lambda}\lambda_{P} g_{\eta\Lambda,b}}{2m_c \xi}\frac{m_{P}+m_{\Lambda}}{\frac{m_{P}^2}{\xi}-m_{\eta_c}^2}\left\{ {\rm exp} \left( -\frac{m_{\eta_c}^2}{T_1^2} \right)-{\rm exp} \left( -\frac{m_{P}^2}{\xi T_1^2} \right) \right\}{\rm exp} \left( -\frac{m_{\Lambda}^2}{T_2^2} \right)\nonumber \\
&& +C_b {\rm exp} \left( -\frac{m_{\eta_c}^2}{T_1^2}-\frac{m_{\Lambda}^2}{T_2^2} \right)=\int_{4m_c^2}^{s_{\eta_c}^0}ds\int_0^{s_{\Lambda}^0}du\, \rho_b(s,u){\rm exp}\left( -\frac{s}{T_1^2}-\frac{u}{T_2^2} \right)\,,
\end{eqnarray}
\begin{eqnarray}
&& \frac{f_{J/\psi}m_{J/\psi}\lambda_{\Lambda}\lambda_{P} }{\xi}\frac{g_{J/\psi\Lambda,T/V}}{\frac{m_{P}^2}{\xi}-m_{J/\psi}^2}\left\{ {\rm exp} \left( -\frac{m_{J/\psi}^2}{T_1^2} \right)-{\rm exp} \left( -\frac{m_{P}^2}{\xi T_1^2} \right) \right\}{\rm exp} \left( -\frac{m_{\Lambda}^2}{T_2^2} \right) \nonumber \\
&& +C_{J/\psi\Lambda,T/V} {\rm exp} \left( -\frac{m_{J/\psi}^2}{T_1^2}-\frac{m_{\Lambda}^2}{T_2^2} \right)\nonumber \\
&&=\int_{4m_c^2}^{s_{J/\psi}^0}ds\int_0^{s_{\Lambda}^0}du\, \rho_{J/\psi\Lambda,T/V}(s,u){\rm exp}\left( -\frac{s}{T_1^2}-\frac{u}{T_2^2} \right)\,,
\end{eqnarray}
\begin{eqnarray}\label{QCDSG-G-i}
&& \frac{f_{\eta_c}m_{\eta_c}^2\lambda_{\Sigma}\lambda_{P^{\prime}} g_{\eta\Sigma,e}}{2m_c \xi}\frac{1}{\frac{m_{P^{\prime}}^2}{\xi}-m_{\eta_c}^2}\left\{ {\rm exp} \left( -\frac{m_{\eta_c}^2}{T_1^2} \right)-{\rm exp} \left( -\frac{m_{P^{\prime}}^2}{\xi T_1^2} \right) \right\}{\rm exp} \left( -\frac{m_{\Sigma}^2}{T_2^2} \right) \nonumber \\
&& +C_e {\rm exp} \left( -\frac{m_{\eta_c}^2}{T_1^2}-\frac{m_{\Sigma}^2}{T_2^2} \right)=\int_{4m_c^2}^{s_{\eta_c}^0}ds\int_0^{s_{\Sigma}^0}du\, \rho_e(s,u){\rm exp}\left( -\frac{s}{T_1^2}-\frac{u}{T_2^2} \right)\,,
\end{eqnarray}
\begin{eqnarray}
&& \frac{f_{\eta_c}m_{\eta_c}^2\lambda_{\Sigma}\lambda_{P^{\prime}} g_{\eta\Sigma,f}}{2m_c \xi}\frac{m_{P^{\prime}}+m_{\Sigma}}{\frac{m_{P^{\prime}}^2}{\xi}-m_{\eta_c}^2}\left\{ {\rm exp} \left( -\frac{m_{\eta_c}^2}{T_1^2} \right)-{\rm exp} \left( -\frac{m_{P^{\prime}}^2}{\xi T_1^2} \right) \right\}{\rm exp} \left( -\frac{m_{\Sigma}^2}{T_2^2} \right)\nonumber \\
&& +C_f {\rm exp} \left( -\frac{m_{\eta_c}^2}{T_1^2}-\frac{m_{\Sigma}^2}{T_2^2} \right)=\int_{4m_c^2}^{s_{\eta_c}^0}ds\int_0^{s_{\Sigma}^0}du\, \rho_f(s,u){\rm exp}\left( -\frac{s}{T_1^2}-\frac{u}{T_2^2} \right)\,,
\end{eqnarray}
\begin{eqnarray}
&& \frac{f_{J/\psi}m_{J/\psi}\lambda_{\Sigma}\lambda_{P^{\prime}} }{\xi}\frac{g_{J/\psi\Sigma,T/V}}{\frac{m_{P^{\prime}}^2}{\xi}-m_{J/\psi}^2}\left\{ {\rm exp} \left( -\frac{m_{J/\psi}^2}{T_1^2} \right)-{\rm exp} \left( -\frac{m_{P^{\prime}}^2}{\xi T_1^2} \right) \right\}{\rm exp} \left( -\frac{m_{\Sigma}^2}{T_2^2} \right) \nonumber \\
&& +C_{J/\psi\Sigma,T/V} {\rm exp} \left( -\frac{m_{J/\psi}^2}{T_1^2}-\frac{m_{\Sigma}^2}{T_2^2} \right)\nonumber \\
&&=\int_{4m_c^2}^{s_{J/\psi}^0}ds\int_0^{s_{\Sigma}^0}du\, \rho_{J/\psi\Sigma,T/V}(s,u){\rm exp}\left( -\frac{s}{T_1^2}-\frac{u}{T_2^2} \right)\,,
\end{eqnarray}
where
\begin{eqnarray}
C_{J/\psi\Lambda,T} &=& \left[ (m_{P}-m_{\Lambda})C_c+C_d \right]\frac{m_{P}+m_{\Lambda}}{m_{P}^2-m_{\Lambda}^2-m_{J/\psi}^2}\,,\nonumber\\
\rho_{J/\psi\Lambda,T}(s,u) &=& \left[ (m_{P}-m_{\Lambda})\rho_c(s,u)+\rho_d(s,u) \right]\frac{m_{P}+m_{\Lambda}}{m_{P}^2-m_{\Lambda}^2-m_{J/\psi}^2}\,,\nonumber\\
C_{J/\psi\Lambda,V} &=& \left( \frac{m_{J/\psi}^2}{m_{P}+m_{\Lambda}}C_c+C_d \right)\frac{m_{P}+m_{\Lambda}}{m_{P}^2-m_{\Lambda}^2-m_{J/\psi}^2}\,,\nonumber\\
\rho_{J/\psi\Lambda,V}(s,u) &=& \left( \frac{m_{J/\psi}^2}{m_P+m_{\Lambda}}\rho_c(s,u)+\rho_d(s,u) \right)\frac{m_{P}+m_{\Lambda}}{m_{P}^2-m_{\Lambda}^2-m_{J/\psi}^2}\,,
\end{eqnarray}
\begin{eqnarray}
C_{J/\psi\Sigma,T} &=& \left[ (m_{P^{\prime}}-m_{\Sigma})C_g+C_h \right]\frac{m_{P^{\prime}}+m_{\Sigma}}{m_{P^{\prime}}^2-m_{\Sigma}^2-m_{J/\psi}^2}\,,\nonumber\\
\rho_{J/\psi\Sigma,T}(s,u) &=& \left[ (m_{P^{\prime}}-m_{\Sigma})\rho_g(s,u)+\rho_h(s,u) \right]\frac{m_{P^{\prime}}+m_{\Sigma}}{m_{P^{\prime}}^2-m_{\Sigma}^2-m_{J/\psi}^2}\,,\nonumber\\
C_{J/\psi\Sigma,V} &=& \left( \frac{m_{J/\psi}^2}{m_{P^{\prime}}+m_{\Sigma}}C_g+C_h \right)\frac{m_{P^{\prime}}+m_{\Sigma}}{m_{P^{\prime}}^2-m_{\Sigma}^2-m_{J/\psi}^2}\,,\nonumber\\
\rho_{J/\psi\Sigma,V}(s,u) &=& \left( \frac{m_{J/\psi}^2}{m_{P^{\prime}}+m_{\Sigma}}\rho_g(s,u)+\rho_h(s,u) \right)\frac{m_{P^{\prime}}+m_{\Sigma}}{m_{P^{\prime}}^2-m_{\Sigma}^2-m_{J/\psi}^2}\,,
\end{eqnarray}
the $T_1$ and $T_2$ are the Borel parameters, the $\rho_Z(s,u)$ are spectral densities at the QCD sides derived from the corresponding correlation functions $\Pi_Z(p'^2,p^2,q^2)$, the subscript $Z$ stands for $a,b,\cdot\cdot\cdot,h$, the $C_Z$ are the unknown parameters determined by choosing flat Borel platforms for the related hadronic coupling constants in the numerical calculations.

\section{QCD sum rules for the $\Lambda$ and $\Sigma$ baryons}
The masses of the $\Lambda$ and $\Sigma$ baryons can be cited from the Particle Data Group \cite{PDG}, however, in order to perform the numerical calculations, we need the pole residues $\lambda_{\Lambda/\Sigma}$ and continuum threshold parameters $s^0_{\Lambda/\Sigma}$, which should be determined by the QCD sum rules. There have been several currents to interpolate the $\Lambda/\Sigma$ baryons, we choose the Ioffe currents \cite{MOka-PRT,ColangeloReview}, and update the calculations. Then, we write down the two-point correlation functions,
\begin{eqnarray}
\Pi_{\Lambda/\Sigma}(q)&=&\textsf{i}\int d^4y e^{\textsf{i}q\cdot y} \langle 0|\texttt{T}\left\{ J_{\Lambda/\Sigma}(y)\bar{J}_{\Lambda/\Sigma}(0)\right\}|0\rangle\, .
\end{eqnarray}
A complete set of baryon states with the same quantum numbers as the currents $J_{\Lambda/\Sigma}(y)$ are inserted into the correlation functions, and only considering  the contributions from the ground states \cite{SVZ1,SVZ2,Reinders}, the hadronic sides are acquired,
\begin{eqnarray}
\Pi_{\Lambda/\Sigma}(q) &=&\lambda_{\Lambda/\Sigma}^2 \frac{\!\not\!{q}+m_{\Lambda/\Sigma}}{m_{\Lambda/\Sigma}^2-q^2}+\cdots\,,\nonumber\\
&=&\Pi_{\Lambda/\Sigma}^1(q^2)\!\not\!{q}+\Pi_{\Lambda/\Sigma}^0(q^2)+\cdots\,.
\end{eqnarray}
The structures $\!\not\!{q}$ and $1$ are chosen, therefore the four  QCD sum rules are  acquired,
\begin{eqnarray}\label{SR-Delta}
\lambda_{\Lambda/\Sigma}^2 {\exp}\left(-\frac{m_{\Lambda/\Sigma}^2}{T^2}\right)&=& \int_0^{s_{\Lambda/\Sigma}^0}\rho^1_{\Lambda/\Sigma,QCD}(u)\,{\rm exp} \left(-\frac{u}{T^2}\right)du\,,\nonumber\\
m_{\Lambda/\Sigma}\lambda_{\Lambda/\Sigma}^2 {\exp}\left(-\frac{m_{\Lambda/\Sigma}^2}{T^2}\right)&=& \int_0^{s_{\Lambda/\Sigma}^0}\rho^0_{\Lambda/\Sigma,QCD}(u)\,{\rm exp} \left(-\frac{u}{T^2}\right)du\,,
\end{eqnarray}
where
\begin{eqnarray}
\rho^{1}_{\Lambda,QCD}(u)&=& \frac{u^2}{64\pi^4}- \left(\frac{1}{3} m_s\langle\bar{q}q\rangle  -\frac{1}{4} m_s \langle\bar{s}s\rangle \right)\frac{1}{\pi^2}+  \langle g_s^2GG\rangle \frac{1}{128\pi^4}\,\nonumber\\
&&+ \left(\frac{1}{12} m_s\langle\bar{q}g_s\sigma Gq\rangle  - \frac{1}{12} m_s \langle\bar{s}g_s\sigma Gs\rangle \right) \frac{\delta(u)}{\pi^2}\,\nonumber\\
&&- \left(\frac{2}{9} \langle\bar{q}q\rangle^2 - \frac{8}{9} \langle\bar{q}q\rangle \langle\bar{s}s\rangle \right) \delta(u) + \frac{1}{27} \langle\bar{q}q\rangle^2 g_s^2  \frac{\delta(u)}{\pi^2}\,\nonumber\\
&&+ \frac{37}{15696} m_s\langle g_s^2GG\rangle \langle\bar{s}s\rangle    \frac{\delta(u)}{\pi^2 T^2}\,\nonumber\\
&&+ \left(\frac{1}{6} \langle\bar{q}q\rangle \langle\bar{q}g_s\sigma Gq\rangle - \frac{2}{9} \langle\bar{q}q\rangle \langle\bar{s}g_s\sigma Gs\rangle  - \frac{1}{9} \langle\bar{q}g_s\sigma Gq\rangle \langle\bar{s}s\rangle \right)  \frac{\delta(u)}{T^2}\,,
\end{eqnarray}
\begin{eqnarray}
\rho^{0}_{\Lambda,QCD}(u) &=& - m_s \frac{u^2}{96 \pi^4}- \left(\frac{1}{3} \langle\bar{q}q\rangle -\frac{1}{12} \langle\bar{s}s\rangle\right) \frac{u}{\pi^2}+ m_s \langle g_s^2 GG \rangle  \frac{1}{384\pi^4}\,\nonumber\\
&&+ \left( \langle\bar{q}g_s\sigma Gq\rangle -  \langle\bar{s}g_s\sigma Gs\rangle \right) \frac{1}{24\pi^2}\,\nonumber\\
&&+ \left(\frac{4}{3} m_s\langle\bar{q}q\rangle^2  - \frac{4}{9} m_s\langle\bar{q}q\rangle \langle\bar{s}s\rangle \right) \delta(u)- m_s \langle\bar{q}q\rangle^2  g_s^2  \frac{\delta(u)}{243\pi^2}\, \,\nonumber\\
&&+ \left(\frac{1}{96} \langle g_s^2GG\rangle \langle\bar{q}q\rangle - \frac{1}{288} \langle g_s^2GG\rangle \langle\bar{s}s\rangle\right)  \frac{\delta(u)}{\pi^2}\,,
\end{eqnarray}
\begin{eqnarray}
\rho^{1}_{\Sigma,QCD}(u)&=&  \frac{u^2}{128 \pi^4}+  m_s \langle\bar{s}s\rangle \frac{1}{8 \pi^2}+  \langle g_s^2 GG \rangle  \frac{1}{256 \pi^4} \,\nonumber\\
&&-  m_s \langle\bar{s}g_s\sigma Gs\rangle   \frac{\delta(u)}{24\pi^2} \,\nonumber\\
&&+ \langle\bar{q}q\rangle^2 \frac{\delta(u)}{3}+\left(\frac{1}{81} \langle\bar{q}q\rangle^2 g_s^2 + \frac{1}{162} \langle\bar{s}s\rangle^2 g_s^2\right) \frac{\delta(u)}{\pi^2}\,\nonumber\\
&&-  \langle\bar{q}q\rangle \langle\bar{q}g_s\sigma Gq\rangle \frac{\delta(u)}{12T^2} \,,
\end{eqnarray}
\begin{eqnarray}
\rho^{0}_{\Sigma,QCD}(u) &=& m_s  \frac{u^2}{64 \pi^4}-\langle\bar{s}s\rangle  \frac{u}{8 \pi^2} -  m_s \langle g_s^2 GG \rangle  \frac{1}{256 \pi^4} \,\nonumber\\
&&+ m_s \langle\bar{q}q\rangle^2 \frac{2\delta(u)}{3} +  m_s g_s^2 \langle\bar{q}q\rangle^2   \frac{\delta(u)}{162\pi^2} \,\nonumber\\
&&+  \langle\bar{s}s\rangle \langle g_s^2 GG \rangle \frac{\delta(u)}{288 \pi^2} \,.
\end{eqnarray}
It is straightforward to obtain the analytical expressions of the masses of the $\Lambda$ and $\Sigma$ baryons,
\begin{eqnarray}\label{Lambda-Sigma-mass}
m^{2}_{\Lambda/\Sigma} &=& \frac{-\frac{\partial}{\partial \tau}\int_0^{s_{\Lambda/\Sigma}^0}\rho^{1/0}_{\Lambda/\Sigma,QCD}(u){\rm exp}\left(-\frac{u}{T^2}\right)du}{\int_0^{s_{\Lambda/\Sigma}^0}\rho^{1/0}_{\Lambda/\Sigma,QCD}(u){\rm exp}\left(-\frac{u}{T^2}\right)du}\,,
\end{eqnarray}
where  $\tau=\frac{1}{T^2}$. The QCD sum rules in Eq.\eqref{Lambda-Sigma-mass} are used to reproduce the experimental values of the masses of the $\Lambda$ and $\Sigma$ baryons \cite{PDG}.
\section{Numerical results and discussions}
For the traditional  QCD sum rules, the vacuum condensates are  input parameters in the numerical calculations, their standard values are determined to be   $\langle\overline{q}q\rangle=-(0.24\pm0.01\,{\rm GeV})^3$,
$\langle\overline{s}s\rangle=(0.8\pm0.1)\langle\overline{q}q\rangle$,
$\langle\overline{q}g_s\sigma Gq\rangle=m_0^2\langle\overline{q}q\rangle$,
$\langle\overline{s}g_s\sigma Gs\rangle=m_0^2\langle\overline{s}s\rangle$,
$m_0^2=(0.8\pm0.1)\,{\rm GeV}^2$, $\langle\frac{\alpha_s}{\pi}GG\rangle=(0.33\,{\rm GeV})^4$ at the energy scale $\mu=1\,{\rm GeV}$ \cite{SVZ1,SVZ2,Reinders,ColangeloReview}, the $\overline{MS}$ masses  $m_{c}(m_c)=(1.275\pm0.025)\,\rm{GeV}$ and $m_s(2\,\rm{GeV})=(0.095\pm 0.005)\,\rm{GeV}$ are taken from the Particle Data Group \cite{PDG}. In the QCD sum rules for the masses and hadronic coupling constants, there are terms of the form $g_s^2\langle \bar{q}q\rangle^2$, which comes from the terms $\langle\bar{q}\gamma_\mu t^aqg_sD_\eta G^{a}_{\lambda\tau}\rangle$, $\langle\bar{q}_jD^{\dag}_\mu D^{\dag}_\nu D^{\dag}_{\alpha} q_i \rangle$ and $\langle \bar{q}_j D_\mu D_\nu D_{\alpha} q_i \rangle$ rather than comes from the perturbative $\mathcal{O}(\alpha_s)$ corrections for the four-quark condensates $\langle \bar{q}q\rangle^2$ \cite{ZGWang-Zc3900-2014,ZGWang-Zb10610-2014,WZG-hidden-cc-PRD}, where $D_\alpha=\partial_\alpha-ig_sG_\alpha $. The strong coupling constant $\alpha_s(\mu)=\frac{g_s^2(\mu)}{4\pi}$ appears at the tree level, which is energy scale dependent, and we should take account of the energy scale dependence in a consistent way. The energy-scale dependence of the input parameters are written as,
\begin{eqnarray}
\langle\overline{q}q\rangle(\mu)&=&\langle\overline{q}q\rangle(1{\rm GeV})\left[\frac{\alpha_s(1{\rm GeV})}{\alpha_s(\mu)}\right]^{\frac{12}{33-2n_f}} \, ,\nonumber\\
\langle\overline{s}s\rangle(\mu)&=&\langle\overline{s}s\rangle(1{\rm GeV})\left[\frac{\alpha_s(1{\rm GeV})}{\alpha_s(\mu)}\right]^{\frac{12}{33-2n_f}} \, ,\nonumber\\
 \langle\overline{q}g_s\sigma Gq\rangle(\mu)& =&\langle\overline{q}g_s\sigma Gq\rangle(1{\rm GeV})\left[\frac{\alpha_s(1{\rm GeV})}{\alpha_s(\mu)}\right]^{\frac{2}{33-2n_f}}\, ,\nonumber\\
 \langle\overline{s}g_s\sigma Gs\rangle(\mu)& =&\langle\overline{s}g_s\sigma Gs\rangle(1{\rm GeV})\left[\frac{\alpha_s(1{\rm GeV})}{\alpha_s(\mu)}\right]^{\frac{2}{33-2n_f}}\, ,\nonumber\\
  m_c(\mu)&=&m_c(m_c)\left[\frac{\alpha_s(\mu)}{\alpha_s(m_c)}\right]^{\frac{12}{33-2n_f}}\, ,\nonumber\\
m_s(\mu)&=&m_s({\rm 2GeV} )\left[\frac{\alpha_{s}(\mu)}{\alpha_{s}({\rm 2GeV})}\right]^{\frac{12}{33-2n_f}}\, ,\nonumber\\ \alpha_s(\mu)&=&\frac{1}{b_0t}\left[1-\frac{b_1}{b_0^2}\frac{\rm{log}\emph{t}}{t}+\frac{b_1^2(\rm{log}^2\emph{t}-\rm{log}\emph {t}-1)+\emph{b}_0\emph{b}_2}{b_0^4t^2}\right]\, ,
\end{eqnarray}
where $t={\rm log}\frac{\mu^2}{\Lambda_{QCD}^2}$, $\emph b_0=\frac{33-2\emph{n}_\emph{f}}{12\pi}$, $b_1=\frac{153-19n_f}{24\pi^2}$, $b_2=\frac{2857-\frac{5033}{9}n_f+\frac{325}{27}n_f^2}{128\pi^3}$
and $\Lambda_{QCD}=213$ MeV, $296$ MeV, $339$ MeV for the flavors $n_f=5,4,3$, respectively \cite{PDG,Narison}, for the strong decays studied in the present work, $n_f=4$, for the QCD sum rules of the $\Lambda$ and $\Sigma$ baryons, $n_f=3$. The energy scales $\mu$ are set as $\frac{m_{\eta_c}}{2}$ for the decay channels $P_{cs}(4338)\rightarrow\eta_c+\Lambda$ and $P_{cs}(4460)\rightarrow\eta_c+\Sigma$, $\mu=\frac{m_{J/\psi}}{2}$ for the decay channels $P_{cs}(4338)\rightarrow J/\psi+\Lambda$ and $P_{cs}(4460)\rightarrow J/\psi+\Sigma$, and the energy scale  is set as $\mu=1\,{\rm GeV}$ for the QCD sum rules of the $\Lambda$ and $\Sigma$ baryons.

For the  mass of the $P_{cs}(4338)$, we use the experimental result $m_{P}=4.338\,{\rm GeV}$ \cite{LHCb-Pcs4338}, and for the mass of the $P_{cs}(4460)$, we follow the conclusion of Ref.\cite{XWWang-Pcs4338-mole}, and  set it  as $m_{P^\prime}=4.460\,{\rm GeV}$. From the Particle Data Group \cite{PDG}, the masses of the mesons and baryons are chosen  as $m_{\eta_c}=2.984\,{\rm GeV}$, $m_{J/\psi}=3.097\,{\rm GeV}$, $m_{\Lambda}=1.116\,{\rm GeV}$ and $m_{\Sigma}=1.189\,{\rm GeV}$, respectively. For the pole residues of the $P_{cs}(4338)$ and $P_{cs}(4460)$, we use the values $\lambda_P=1.43\times10^{-3}\,{\rm GeV^6}$ and $\lambda_{P'}=1.37\times10^{-3}\,{\rm GeV^6}$ in Ref.\cite{XWWang-Pcs4338-mole}. For the decay constants, we take the values  $f_{J/\psi}=0.418\,{\rm GeV}$ and $f_{\eta_c}=0.387\,{\rm GeV}$ from the QCD sum rules combined with lattice QCD \cite{Becirevic}.
As for the continuum threshold parameters $s_{\eta_c}^0$ and $s_{J/\psi}^0$, we choose the values $\sqrt{s_{\eta_c}^0}=3.50\,{\rm GeV}$ and $\sqrt{s_{J/\psi}^0}=3.60\,{\rm GeV}$ \cite{Decay-mole-WZG-WX}.

For the QCD sum rules of the $\Lambda$ and $\Sigma$ baryons, the masses and pole residues are phenomenologically solved via the averages  $m_{\Lambda/\Sigma}=\frac{1}{2}(m_{\Lambda/\Sigma}^1+m_{\Lambda/\Sigma}^0)$ and $\lambda_{\Lambda/\Sigma}=\frac{1}{2}(\lambda_{\Lambda/\Sigma}^1+\lambda_{\Lambda/\Sigma}^0)$, where the $m_{\Lambda/\Sigma}^{1/0}$ and $\lambda_{\Lambda/\Sigma}^{1/0}$ are  derived from the spectral densities $\rho_{\Lambda/\Sigma,QCD}^{1/0}(u)$. The diagrams $m_{\Lambda}-T^2$ and $\lambda_{\Lambda}-T^2$ are shown in the Fig.1, setting  $\sqrt{s_{\Lambda}^0}=1.59\,{\rm GeV}$, the Borel window  ranges from $T^2_{min}=1.10\,{\rm GeV^2}$ to $T^2_{max}=1.50\,{\rm GeV^2}$ with the pole contribution being $(43-62)\%$,  the central value of the extracted mass is $1.118\,{\rm{GeV}}$, which coincides  with the experimental data $m_{\Lambda}=1.116\,{\rm{GeV}}$ from the Particle Data Group \cite{PDG}, then the pole residue is determined as $\lambda_{\Lambda}=2.87\times 10^{-2}\,{\rm GeV^3}$. The diagrams $m_{\Sigma}-T^2$ and $\lambda_{\Sigma}-T^2$ are shown in the Fig.2, if we take $\sqrt{s_{\Sigma}^0}=1.68\,{\rm GeV}$, the Borel window ranges from $T^2_{min}=1.15\,{\rm GeV^2}$ to $T^2_{max}=1.55\,{\rm GeV^2}$, then the pole contribution  is $(41-61)\%$, and the central value of the extracted mass  is $1.188\,{\rm{GeV}}$, which  coincides with the experimental data $m_{\Sigma}=1.189\,{\rm{GeV}}$ from the Particle Data Group \cite{PDG},  the pole residue is determined  as $\lambda_{\Sigma}=2.17\times 10^{-2}\,{\rm GeV^3}$.

For the spectral densities $\rho_{Z}$, $Z=a,b,\cdot\cdot\cdot,h$, they all contain two Borel parameters $T_1^2$ and $T_2^2$, namely, $\rho_{Z}=\rho_{Z}(T_1^2,T_2^2)$, obviously, the hadronic coupling constants rely on $T_1^2$ and $T_2^2$, that is, $g=g(T_1^2,T_2^2)$. For the QCD sum rules, the error bounds due to the Borel parameters should be small, thus, flat Borel platform should be obtained to extract the physical quantities, then, it is straightforward to set $T_1^2=T_2^2=T^2$. Via trivial and error, the Borel platforms of each hadronic coupling constants are obtained  by setting the free parameters as $C_a=5.95\times 10^{-6}\,{\rm GeV}^9$, $C_b=3.86\times 10^{-5}\,{\rm GeV}^{10}$, $C_{J/\psi\Lambda,T}=1.80\times 10^{-5}\,{\rm GeV}^{11}$, $C_{J/\psi\Lambda,V}=2.50\times 10^{-5}\,{\rm GeV}^{11}$, $C_e=-1.23\times 10^{-5}\,{\rm GeV}^9+4.92\times 10^{-7}T^2\,{\rm GeV}^7$, $C_f=-7.52\times 10^{-5}\,{\rm GeV}^{10}+3.01\times 10^{-7}T^2\,{\rm GeV}^8$, $C_{J/\psi\Sigma,T}=-2.87\times 10^{-5}\,{\rm GeV}^{11}+1.18\times 10^{-7}T^2\,{\rm GeV}^9$, $C_{J/\psi\Sigma,V}=-3.67\times 10^{-5}\,{\rm GeV}^{11}-2.93\times 10^{-8}T^2\,{\rm GeV}^9$.

In order to estimate the error bounds, the approximations $\frac{\delta \lambda_P}{\lambda_P}=\frac{\delta \lambda_{P^\prime}}{\lambda_{P^\prime}}=\frac{\delta \lambda_{\Lambda}}{\lambda_{\Lambda}}=\frac{\delta \lambda_{\Sigma}}{\lambda_{\Sigma}}=\frac{\delta f_{J/\psi}}{f_{J/\psi}}=\frac{\delta f_{\eta_c}}{f_{\eta_c}}$ are applied \cite{WZG-Pc4312-decay-tetra,DZGWang,DZGWang-2}, moreover, the error bounds due to the parameters $C_Z$ are not considered, under such considerations, the $g-T^2$ graphs are shown in the Figs.3-6, and their numerical results are extracted as,
\begin{eqnarray}
g_{\eta\Lambda,a}=0.184^{+0.048}_{-0.048}\, ,&&T^2=5.5-6.5\,{\rm GeV}^2\, ,\nonumber\\
g_{\eta\Lambda,b}=0.181^{+0.054}_{-0.054}\, ,&&T^2=5.5-6.5\,{\rm GeV}^2\, ,\nonumber\\
g_{J/\psi\Lambda,T}=0.135^{+0.042}_{-0.042}\, ,&&T^2=5.5-6.5\,{\rm GeV}^2\, ,\nonumber\\
g_{J/\psi\Lambda,V}=0.371^{+0.107}_{-0.107}\, ,&&T^2=6.0-7.0\,{\rm GeV}^2\, ,\nonumber\\
g_{\eta\Sigma,e}=0.577^{+0.096}_{-0.096}\, ,&&T^2=5.5-6.5\,{\rm GeV}^2\, ,\nonumber\\
g_{\eta\Sigma,f}=0.577^{+0.106}_{-0.106}\, ,&&T^2=6.0-7.0\,{\rm GeV}^2\, ,\nonumber\\
g_{J/\psi\Sigma,T}=1.012^{+0.215}_{-0.215}\, ,&&T^2=6.0-7.0\,{\rm GeV}^2\, ,\nonumber\\
g_{J/\psi\Sigma,V}=0.129^{+0.314}_{-0.324}\, ,&&T^2=6.0-7.0\,{\rm GeV}^2\, .
\end{eqnarray}
Based on the hadronic coupling constants, we obtain the corresponding  partial decay widths directly,
\begin{eqnarray} \label{Par-width}
\Gamma^{a} (P_{cs}(4338)\rightarrow \eta_c \Lambda) &=& 0.95^{+0.50}_{-0.50} \,{\rm MeV}\,, \nonumber\\
\Gamma^{b} (P_{cs}(4338)\rightarrow \eta_c \Lambda) &=& 0.92^{+0.55}_{-0.55} \,{\rm MeV}\,, \nonumber\\
\Gamma (P_{cs}(4338)\rightarrow J/\psi \Lambda) &=& 5.21^{+8.00}_{-5.21} \,{\rm MeV}\,, \nonumber\\
\Gamma^{e} (P_{cs}(4460)\rightarrow \eta_c \Sigma) &=& 11.04^{+3.67}_{-3.67} \,{\rm MeV}\,, \nonumber\\
\Gamma^{f} (P_{cs}(4460)\rightarrow \eta_c \Sigma) &=& 11.04^{+4.06}_{-4.06} \,{\rm MeV}\,, \nonumber\\
\Gamma (P_{cs}(4460)\rightarrow J/\psi \Sigma) &=& 14.76^{+22.71}_{-14.76} \,{\rm MeV}\,,
\end{eqnarray}
where the $\Gamma^{a/b}$ are due to the hadronic coupling constants $g_{\eta\Lambda,a/b}$, respectively,  the $\Gamma^{e/f}$ are due to the hadronic coupling constants  $g_{\eta\Sigma,e/f}$, respectively. Taking the average value $\frac{1}{2}(\Gamma^{a}+\Gamma^{b})$, the width of the $P_{cs}(4338)$ is then $6.15\,{\rm MeV}$, it is in good agreement with experimental data \cite{LHCb-Pcs4338}, moreover, the ratio of the partial decay widths,
\begin{eqnarray}
\frac{\Gamma(P_{cs}(4338)\rightarrow \eta_c \Lambda)}{\Gamma(P_{cs}(4338)\rightarrow J/\psi \Lambda)}&=&0.18\, .
\end{eqnarray}
As for the $P_{cs}(4460)$, which is the high isospin cousin of the $P_{cs}(4338)$, we take the average value $\frac{1}{2}(\Gamma^{e}+\Gamma^{f})$, the width of this resonance state is then $25.80\,{\rm MeV}$, the ratio of its partial decay widths,
\begin{eqnarray}
\frac{\Gamma(P_{cs}(4460)\rightarrow \eta_c \Sigma)}{\Gamma(P_{cs}(4460)\rightarrow J/\psi \Sigma)}&=&0.48\,.
\end{eqnarray}
\section{Conclusions}
In the present work, the   hadronic coupling constants in the two-body strong decays of the $P_{cs}(4338)$ and $P_{cs}(4460)$ are studied via the QCD sum rules. The numerical results show that the theoretical  calculations are in good agreement with the experimental data for the width of the $P_{cs}(4338)$, the ratio of the partial decay widths is waiting for more experimental data to  testify, it will in return judge our interpretation of the physical nature of the exotic pentaquark candidate $P_{cs}(4338)$. The hadronic decay widths for the predicted state $P_{cs}(4460)$ are also obtained, which would present additional information for the possible observation of this state in the future experiment.

\section*{Acknowledgements}
This work is supported by National Natural Science Foundation, Grant Number 12175068 and the Fundamental Research Funds for the Central Universities of China.

\clearpage
\begin{figure}
 \centering
 \includegraphics[totalheight=5cm,width=7cm]{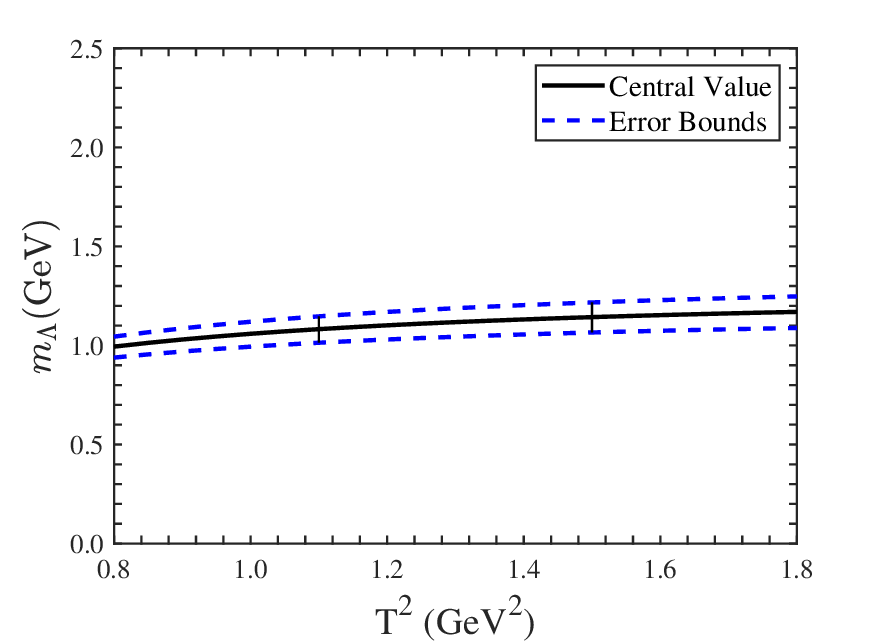}
 \includegraphics[totalheight=5cm,width=7cm]{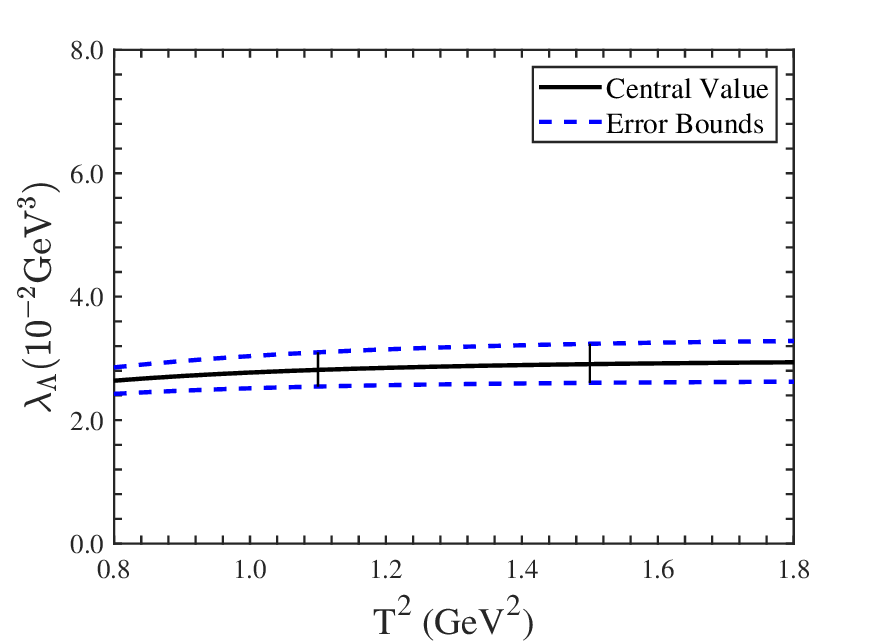}
 \caption{The numerical results of the mass (Left) and pole residue (Right) of the $\Lambda$ baryon.}\label{baryon1-fig}
\end{figure}
\begin{figure}
 \centering
 \includegraphics[totalheight=5cm,width=7cm]{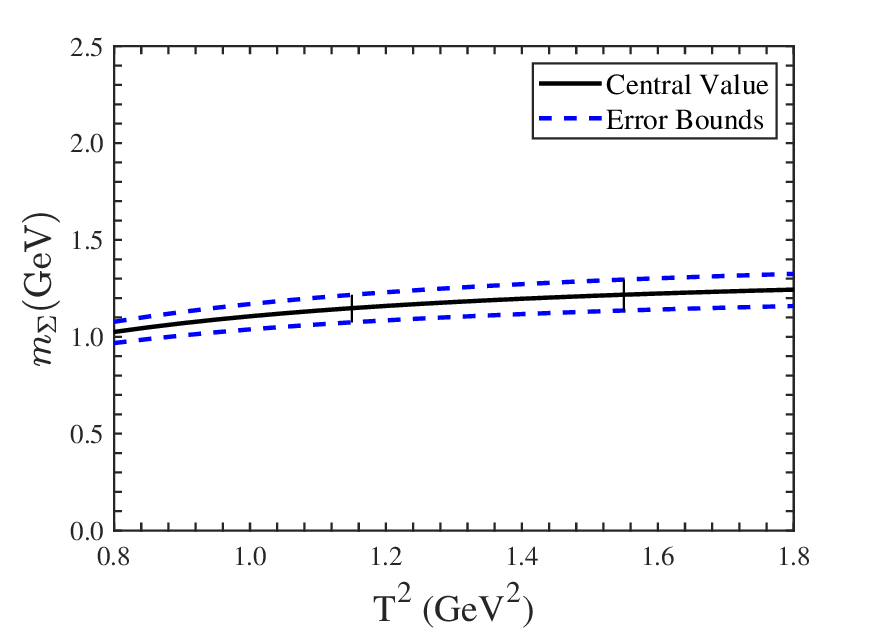}
 \includegraphics[totalheight=5cm,width=7cm]{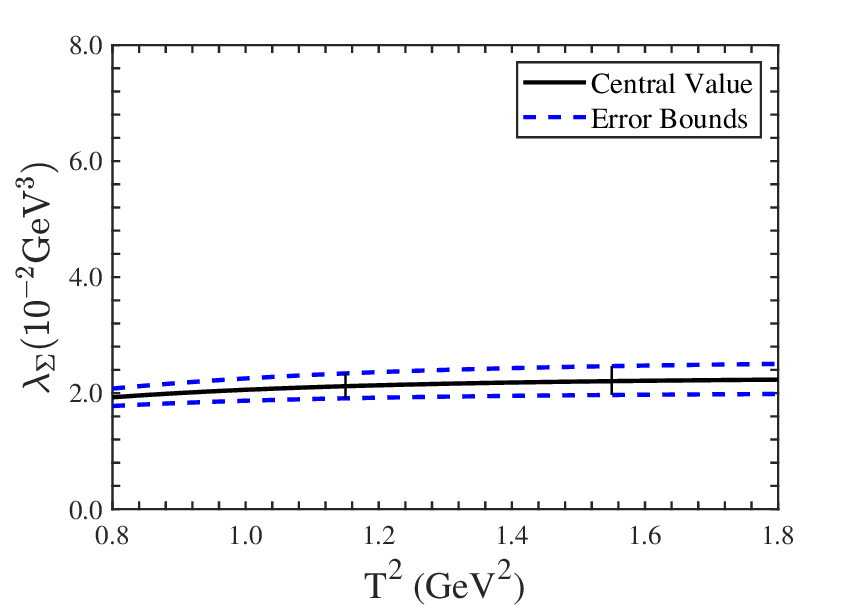}
 \caption{The numerical results of the mass (Left) and pole residue (Right) of the $\Sigma$ baryon.}\label{baryon2-fig}
\end{figure}
\begin{figure}
 \centering
 \includegraphics[totalheight=5cm,width=7cm]{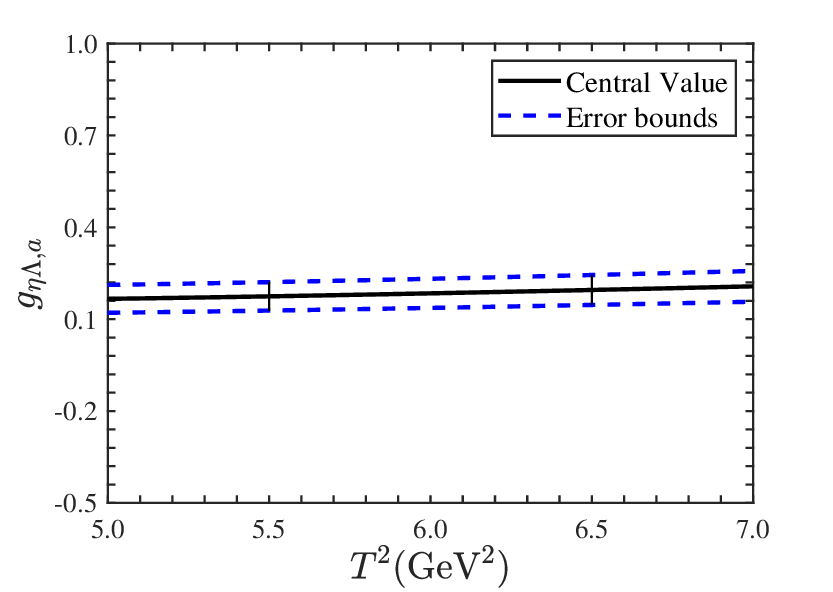}
  \includegraphics[totalheight=5cm,width=7cm]{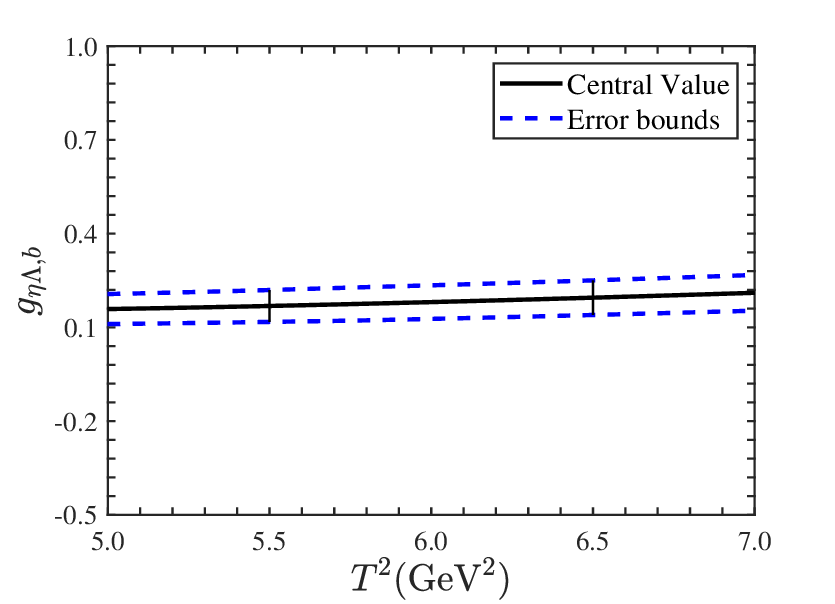}
 \caption{The $g_{\eta\Lambda,a}-T^2$ (Left) and $g_{\eta\Lambda,b}-T^2$ (Right) curves, where the region  among the two short vertical lines of each graph represents the Borel platforms.}\label{3Dfigg1a}
\end{figure}
\clearpage
\begin{figure}
 \centering
 \includegraphics[totalheight=5cm,width=7cm]{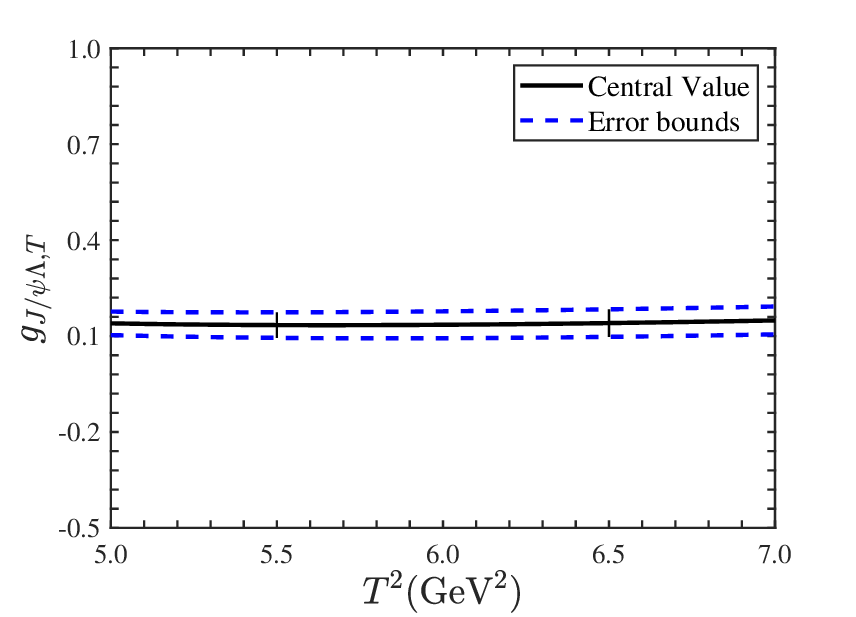}
  \includegraphics[totalheight=5cm,width=7cm]{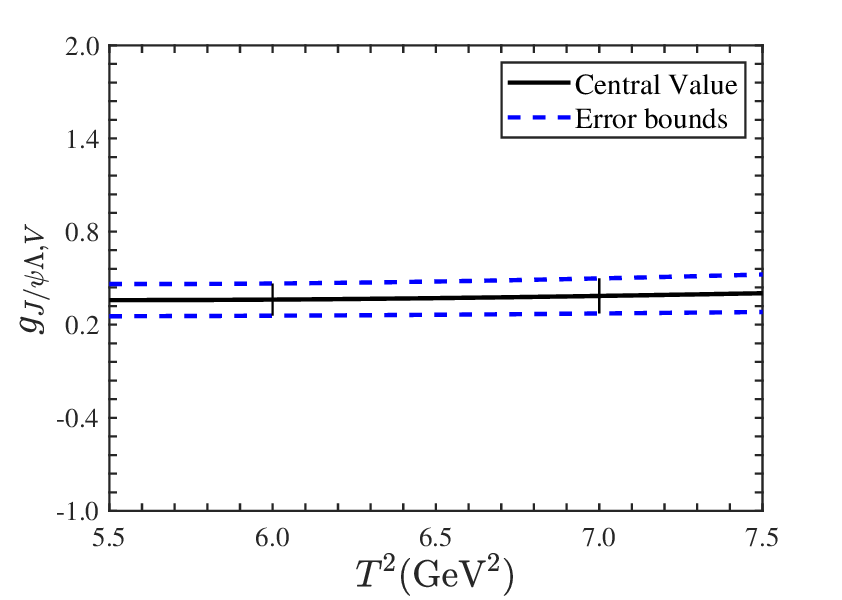}
 \caption{The $g_{J/\psi\Lambda,T}-T^2$ (Left) and $g_{J/\psi\Lambda,V}-T^2$ (Right) curves, where the region  among the two short vertical lines of each graph represents the Borel platforms.}\label{3Dfigg1a}
\end{figure}
\begin{figure}
 \centering
 \includegraphics[totalheight=5cm,width=7cm]{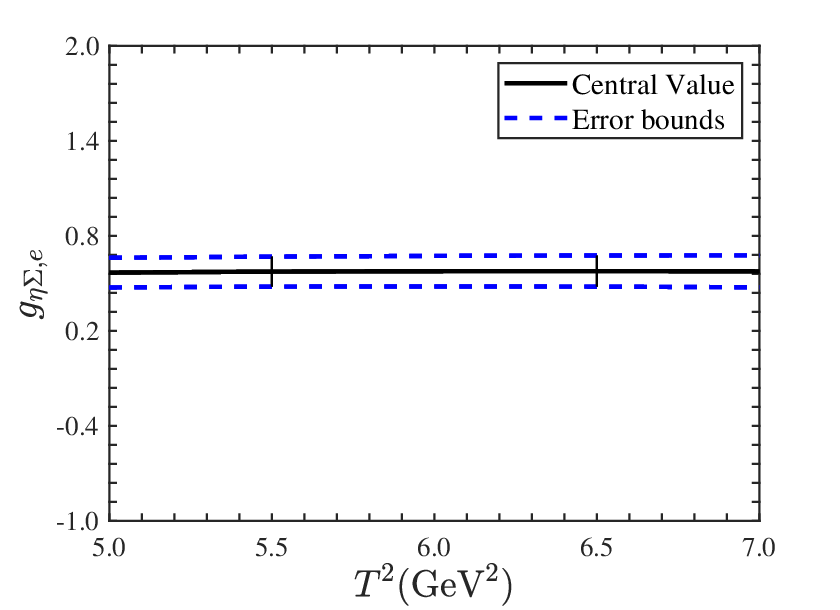}
  \includegraphics[totalheight=5cm,width=7cm]{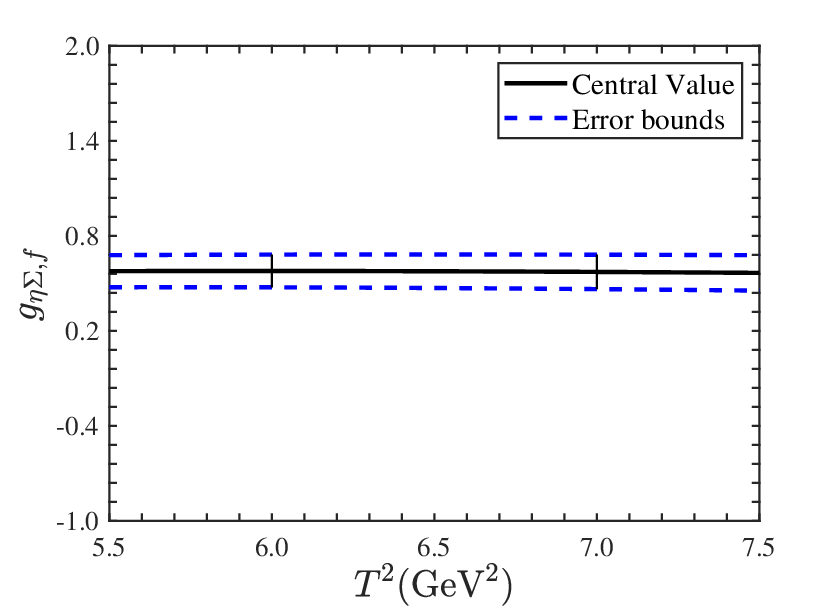}
 \caption{The $g_{\eta\Sigma,e}-T^2$ (Left) and $g_{\eta\Sigma,f}-T^2$ (Right) curves, where the region among the two short vertical lines of each graph represents the Borel platforms.}\label{3Dfigg1a}
\end{figure}
\begin{figure}
 \centering
 \includegraphics[totalheight=5cm,width=7cm]{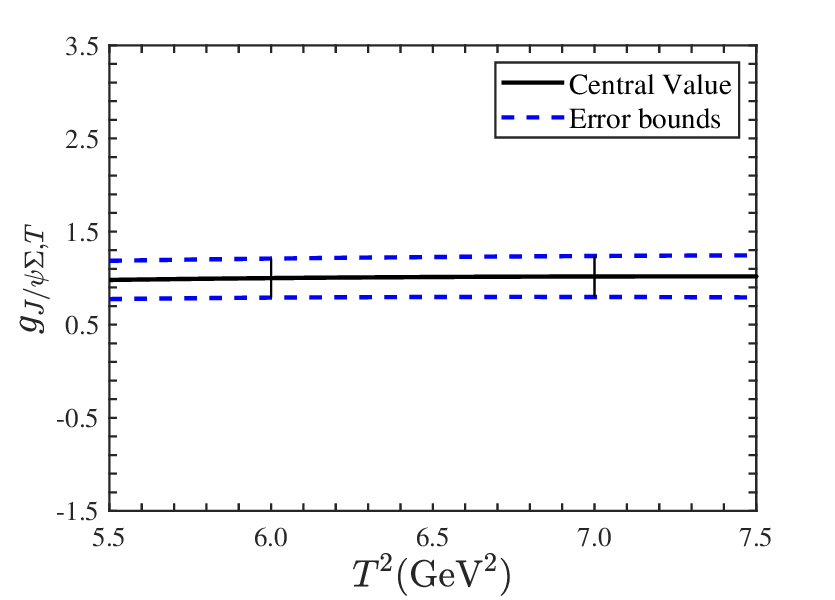}
  \includegraphics[totalheight=5cm,width=7cm]{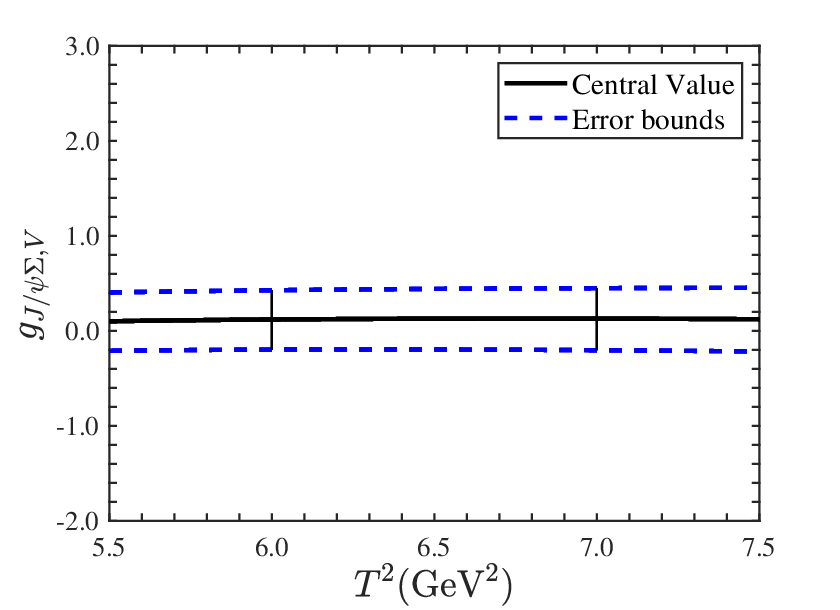}
 \caption{The $g_{J/\psi\Sigma,T}-T^2$ (Left) and $g_{J/\psi\Sigma,V}-T^2$ (Right) curves, where the region among the two short vertical lines  of each graph represents the Borel platforms.}\label{3Dfigg1a}
\end{figure}
\clearpage

\end{document}